# Examining the tech stacks of Czech and Slovak untrustworthy websites

Jozef Michal Mintal[1;2;3] [0000-0003-1651-7146], Anna Macko[2], Marko Paľa[2], Františka Pirosková[2], Pavlo Yakubets[2], Jaroslav Ušiak[1;2] [0000-0001-9375-2807], and Karol Fabián[1;2] [0000-0003-1424-0737]

[1] Department of Security Studies, Matej Bel University, Banská Bystrica, Slovakia
`jozef.mintal@umb.sk`
[2] UMB Data&Society Lab, CEKR, Matej Bel University, Banská Bystrica, Slovakia
[3] Center for Media, Data and Society at CEU Democracy Institute, Central European University, Budapest, Hungary

**Abstract.** The burgeoning of misleading or false information spread by untrustworthy websites has, without doubt, created a dangerous concoction. Thus, it is not a surprise that the threat posed by untrustworthy websites has emerged as a central concern on the public agenda in many countries, including Czechia and Slovakia. However, combating this harmful phenomenon has proven to be difficult, with approaches primarily focusing on tackling consequences instead of prevention, as websites are routinely seen as quasi-sovereign organisms. Websites, however, rely upon a host of service providers, which, in a way, hold substantial power over them. Notwithstanding the apparent power hold by such tech stack layers, scholarship on this topic remains largely limited. This article contributes to this small body of knowledge by providing a first-of-its-kind systematic mapping of the back-end infrastructural support that makes up the tech stacks of Czech and Slovak untrustworthy websites. Our approach is based on collecting and analyzing data on top-level domain operators, domain name Registrars, e-mail providers, web hosting providers, and utilized website tracking technologies of 150 Czech and Slovak untrustworthy websites. Our findings show that the Czech and Slovak untrustworthy website landscape relies on a vast number of back-end services spread across multiple countries, but in key tech stack layers is nevertheless still heavily dominated by locally based companies. Finally, given our findings, we discuss various possible avenues of utilizing the numeral tech stack layers in combating online disinformation.

## 1 Introduction

The proliferation of misleading or false information in cyberspace has emerged as a significant concern on the public agenda in recent years [1, 2]. Around the globe, a variety of politically and economically motivated actors have been shown to deliberately spread lies online [3]. These actors tend to employ a host of tactics to achieve their desired ends. A common technique is the creation of a disinformation ecology consisting of one or multiple websites producing false or misleading content that afterward gets amplified by interconnected social networking service accounts [4]. Such disinformation ecologies can be observed in multiple languages, including Czech and Slovak. Effectively combating this harmful phenomenon has, however, from a governance perspective proven to be difficult.

Policy response tackling the issues of disinformation, as shown by Tenove [2], has usually fallen in three governance sectors: international and national security policies, electoral regulation, and media regulation, with the last garnering much of the public and scholarly attention. Such publicity is likely also fueled by a sizeable distaste aimed at social networking services (SNSs) such as Facebook or Twitter, which have been accused of enacting lax content moderation policies combined with limited or no enforcement. Thus, it might not come as a surprise that the debate about media regulation tends to focus on the 'community standards' of major social networking services or around individual website policies [5]. Even though much important, such a debate approaches content governance from a too narrow a view, missing the overall digital governance ecosystem picture along the way. Social networking services, as well as individual websites, rely upon, and in multiple instances, are directly dependent on a host of service providers, who, in a way, hold substantial power to compel SNSs and websites to adopt more strict content moderation. When approaching the issue of moderating misleading or false information, from such a tech stack perspective, a multitude of additional regulation strategies become apparent. As Donovan [5] has argued, a tech stack approach to content moderation comprises several levels, including domain registrars, cloud services, and internet service providers. Power of tech stacks can be observed in multiple instances, such as EURid's decision to amend its Abuse Prevention and Early Warning System, as to ramp up the blocking of nefarious registration of covid related domain names [6], or Amazon's recent decision to suspend web hosting for an entire SNS called Parler, following its utilization in the 2021 storming of the U.S. Capitol [7].

Notwithstanding the apparent power hold by tech stacks, scholarship on this topic in the context of countering untrustworthy websites remains to a great extend limited [5, 8], with empirical studies being largely confined to one recent paper by Au, Howard and Ainita [8]. Thus, the aim of this article is to (i) provide a mapping of the back-end services used to sustain untrustworthy websites, piloted in Czechia and Slovakia; and (ii) to explore possible avenues of utilizing the identified tech stacks of untrustworthy websites in curbing out the threat of online disinformation. To achieve the above aim, the presented article first offers background information on untrustworthy websites in Czechia and Slovakia. Then, it introduces the methods and results of our empirical study. Finally, based on the findings of our study, we explore in this article some of the possible measures that might help in combating the dangerous phenomenon of online disinformation.

## *1.1* The Czech and Slovak untrustworthy website landscape

In recent years, untrustworthy websites [9] started burgeoning across many countries around the globe, with Czechia and Slovakia being no exception. As Klingová has noted, the exact number of such websites operating natively in these two countries is difficult to pinpoint, as new websites keep on getting established while some of the old ones cease to exist [10]. Albeit estimates typically range in the couple hundreds, one of the most comprehensive publicly available list of relevant Slovak and Czech untrustworthy websites — konspiratori.sk, which is heavily relied upon by local researchers and policy experts [10, 11], lists 205 websites as of February 2021 [12]. Thematically the websites tracked in the konspiratori.sk list cover a wide range of topics, including, among others, health disinformation, Russian propaganda and the paranormal. Even though some of these untrustworthy websites boast a high number of monthly visitors, with a couple of them even making it into the top 50 most visited websites in their respective countries, transparency wise a sizeable amount of these online outlets remain shrouded in mystery [11, 13, 14]. As Mintal and Rusnak have noted, from an operational and financial perspective, at least in Slovakia, it appears that the untrustworthy website landscape is run by multiple independent entities using various business models to sustain operation[1] [11]. The blend of harmful and potentially also illegal content spread by, in some instances, highly viral untrustworthy websites has, without doubt, created a dangerous concoction. Thus, it might not come as a surprise that the threat posed by untrustworthy websites has emerged as a central concern on the public agenda in both Czechia and Slovakia.

## 2 Methods

To study the possible tech-stack moderation areas of Czech and Slovak untrustworthy websites, we used a variety of open-source and proprietary tools to collect data and run analyses. Our overall methodology builds on Donovan's [5] concept of a tech stack moderation approach to online content and Au et al.'s empirical study of infrastructural support for controversial covid-19 websites [8]. Stemming from these approaches, in our study, we examined Czech and Slovak untrustworthy websites' (i) top-level domain operators, (ii) domain name registrars, (iii) e-mail providers, (iv) web hosting providers, and utilized (v) website tracking technologies. Data Collection was conducted during the second half of February 2021 and consisted of multiple interlinked steps. First, for our sample size, we used a dataset of 205 untrustworthy websites, taken from the konspiratori.sk database [12]. Inclusion of a website in the database is assessed based on a set of publicly available criteria by a Review Board consisting of, among others, prominent historians, political scientists, medical professionals, journalists and civil society representatives [15]. In utilizing the konspiratori.sk database, we followed best practices used in Czech and Slovak disinformation scholarship [10, 11]. Second, each website in our dataset was manually checked to determine its availability status and primary language. Unavailable websites and websites with a primary language other than Czech or Slovak were discarded. The remaining websites were then manually checked for affiliated social networking service accounts and manually content-coded according to a slightly adapted framework by Mintal and Rusnak [11]. Third, we used two dedicated services to query domain information and DNS records of the websites under investigation [16, 17]. Data collection was in this step focused mainly on identifying the domain registrars, hosting providers and e-mail providers. In addition, we also employed a bulk domain to IP lookup tool to cross-check the previously collected information on hosting providers [18]. Fourth, to detect web tracking technologies running on the Czech and Slovak untrustworthy websites in our dataset, we used the DMI App Tracker script and cross-checked the data with data obtained from Whatruns [19, 20]. The collected web tracking technology data was then coded according to the DMI Tracker framework [19]. Furthermore, we checked all URLs contained in our untrustworthy websites dataset against the top one-hundred most visited websites in Czechia and Slovakia according to an enterprise-grade web analytics firm [21].

---

[1] Popular income sources of untrustworthy websites in Slovakia include tax designation, e-commerce, crowdfunding, and advertising [11].

After the data collection phase ended, all generated datasets underwent data cleansing to detect and correct corrupt or inaccurate records; duplicate data was discarded. Subsequently, we calculated basic descriptive statistics for the cleansed datasets and identified the most salient organizations providing infrastructural support to the untrustworthy websites under investigation. Last, we formatted the resultant datasets and visualized the data.

## 3 Results

Out of our initial sample size of 205 untrustworthy websites, 49 appeared to cease to exist and six were found to be primarily in languages other than Czech or Slovak, and therefore discarded from our dataset. Thus, our final list of untrustworthy websites consisted of 150 websites, out of which 69 were primarily in Slovak and 81 were primarily in Czech. Manual content-coding of the final list, utilizing a modified framework by Mintal and Rusnak [11] showed that the majority of websites under investigation were News-Focused with 72 entries, followed by the Ideological or Supporting Cause category with 49 entries, Health and Lifestyle category with 20 entries, and the Paranormal category with nine entries.

### 3.1 Top-level domain names and domain name registrars

Querying the tech stack of the websites included in our final list showed that ten of them were hosted on a microblogging or social networking service such as tumblr.com or livejournal.com. To avoid skewed results, in instances where this might have affected the results of our empirical study, such as in mapping the domain registrars or hosting providers, we split the dataset into two — one containing the ten microblogging or social networking service hosted websites and the other dataset containing the rest of the websites under investigation. An analysis of top-level domains (TLDs) and registrars used by the Czech and Slovak untrustworthy websites revealed that the majority of these online outlets tend to use Czech or Slovak TLDs and registrars (see Fig. 1). However, a smaller number of websites use third country services for TLDs and registrars, with some of them utilizing highly privacy and free speech friendly providers in countries such as the United States of America, Panama or the Bahamas.

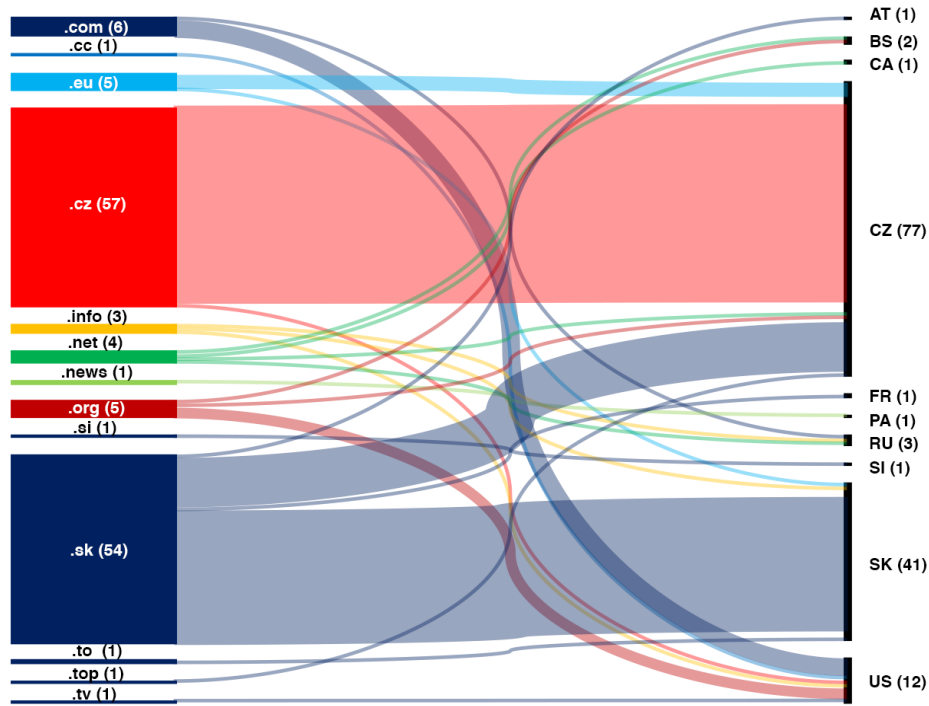

**Fig. 1.** Top-level domain name suffixes of Czech and Slovak untrustworthy websites by the corresponding country code of the utilized domain name registrar.

### 3.2 Hosting and e-mail providers

The Czech and Slovak untrustworthy website landscape under investigation utilizes a vast pool of geographically dispersed hosting providers spanning eleven countries. Among these locations, the strongest represented ones are Czechia with 60 entries, followed by Slovakia with 29 entries, the United States of America with 29 entries, and Germany with 14 entries. As for the most utilized hosting providers, two companies are tied for the first place, with each having 13 entries — namely the Czech-based Wedos Internet a.s., and Slovak-based Websupport s. r. o. Apart from the geographical dispersity of the utilized hosting providers, it appears that a sizeable proportion of websites use a content delivery network (CDN), with the most popular being U.S.-based Cloudflare with 13 entries. In regard to e-mail providers, the websites under investigation almost exclusively relied upon an e-mail service provided to them by their hosting provider, with a few exceptions in which websites relied upon a third-party e-mail provider such as Gmail or the privacy-friendly Protonmail.

### 3.3 Trackers

In our analysis, we detected web tracking technologies on 126 websites under investigation. The most popular trackers present in our dataset were DoubleClick, followed by Google Analytics, and Facebook Connect (see Fig. 2).

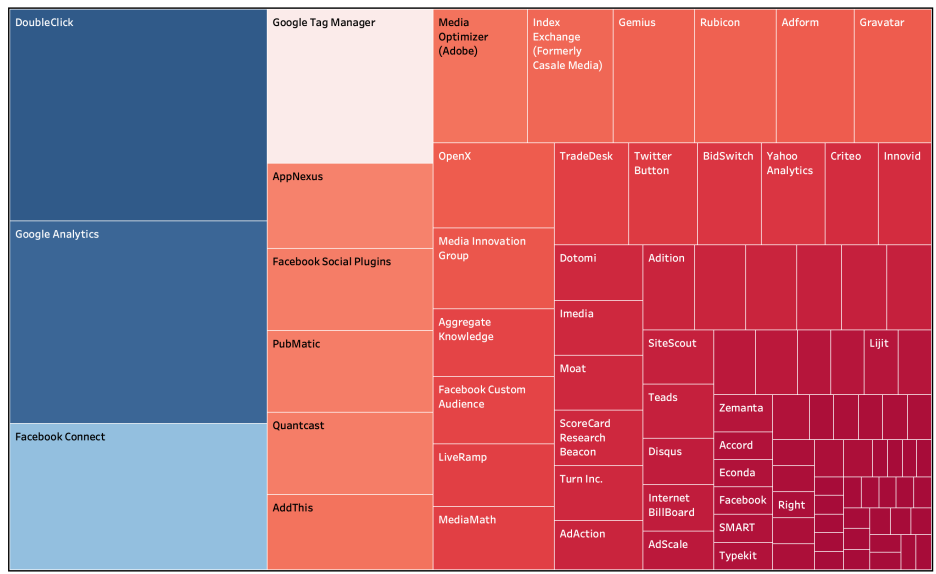

**Fig. 2.** Treemap of trackers present on Czech and Slovak untrustworthy websites

In aggregate, we observed 949 trackers with 93 different fingerprints of tracking technologies being represented. Coding of the observed trackers utilizing the DMI Tracker framework [19] revealed that 49.10% were advertising elements, 23.07% widget elements, 17.28% analytics elements and 10.53% tracker elements.

## 4 Discussion

The results described above show that the Czech and Slovak untrustworthy website landscape relies on a vast number of back-end services spread across multiple countries. Even though this geographical diversity is eminent, it can be argued that core layers of these websites' tech stacks, in areas such as TLD operators, domain registrars, and even hosting providers, are nevertheless still heavily dominated by Czech and Slovak based companies.

So, what does this situation mean in terms of utilizing the identified tech stacks of untrustworthy websites to chart possible avenues in combating online disinformation? For the two local governments in question, the reliance of untrustworthy websites on local companies, means among other things, direct jurisdiction over critical areas of the untrustworthy websites' tech stacks. Such control theoretically opens up the possibility of combating the local untrustworthy website landscape through regulation affecting technological intermediaries such as hosting providers or TLD operators. However, an approach like this would likely be seen as highly controversial as, for one, spreading online disinformation might not necessarily be illegal. Besides the fact that not all disinformation can be considered illegal, some critics point out to a supposed nonequivalence between online and offline law, calling for enacting new harmonizing legislative measures. However, ultimately such calls might, among other things, be a red herring. Various legal experts have noted, that generally, behaviour is broadly criminalized to the same extent online as offline [22]; what is, however, oftentimes lacking in regard to the online domain are case law, human and technical capital resources [23] and overall a more active pursue in terms of criminal law enforcement [22]. This line of argument can be illustrated on the Slovak case of the untrustworthy website badatel.net, which, even though has for many years spread disinformation that might be considered illegal, only recently, in 2020, a major legal action has been taken against this website [24]. A somewhat similar situation can be seen in Czechia, where local law enforcement have in the wake of the covid-19 pandemic, more actively engaged in investigating long time disinformation spreading websites such as otevrisvoumysl.cz or ac24.cz [25].

However, governments and law enforcement do not have to be the only driving force behind combating untrustworthy websites. Various key service providers present in the tech stacks of Czech and Slovak untrustworthy websites already have provisions in their Terms of Services that might be used to combat harmful disinformation. A case in point is the action taken by EURid, which is the registry manager of the .eu top-level domain name. As a result of the covid-19 pandemic, EURid has amended its already used Abuse Prevention and Early Warning System, as to ramp up the blocking of nefarious registrations of covid-19 related domain names [6]. However, as a recent report by the Digital Citizens Alliance has found out, multiple major domain Registrars have, in general, done little to prevent the sale of domain names with the potential to spread covid-19 related disinformation [26].

The results of this study, however, bring forth another possible avenue of combating untrustworthy websites. Even though the various layers of a tech stack are from a technical point of view not equally important in keeping a website online, some less critically important services might theoretically wield great indirect power over the actual operational existence of a website. As demonstrated by our data on website tracking technologies present on Czech and Slovak untrustworthy websites, the most widely used tracking elements were directly related to advertising. This finding of heavy reliance on ad serving services is also in line with previous studies exploring the business models of Slovak [11] and Czech [13] untrustworthy websites. Thus, a decision by a major internet ad serving service such as DoubleClick or PubMatic to demonetize untrustworthy websites could have potentially far-reaching implications for the continued existence of such websites and whole disinformation ecologies, thus consequently providing an indirect tool of combating online disinformation. Nevertheless, as with the other in this article discussed avenues of utilizing the power of tech stacks to combat disinformation, the demonetization approach as well, is in some ways vulnerable to the possibility that untrustworthy website owners could create new websites and, in some instances, even shift their already established ones, to rely on tech stack layers based in privacy and free speech friendly jurisdictions. Nonetheless, a demonetization approach might be highly effective at targeting well-established high web traffic untrustworthy websites.

In conclusion, besides providing a first of its kind systematic mapping of the back-end infrastructural support that makes up the tech stacks of Czech and Slovak untrustworthy websites, our article highlighted some of the possible avenues of utilizing the numeral tech stack layers in combating online disinformation. Domain name providers, ad serving services, hosting providers and other various back-end infrastructural services sustaining untrustworthy websites can all play a critical role in creating a safer infosphere. Nevertheless, it must be noted that to achieve such a goal, there is no one silver bullet, instead what appears to be needed is a coordinated approach among various stakeholders, potentially on an international level.

**Funding and declaration of conflicting interest.**
This work was supported by Erasmus+ Strategic Partnerships for higher education (grant number 2020-1-IT02-KA203-079902). The authors declared no potential conflicts of interest with respect to the research, authorship, and/or publication of this article.

**Contributions**
J.M.M. designed the study and supervised the research; A.M., M.P., F.P., P.Y. collected the data with input from J.M.M.; J.U. provided the funding; J.M.M. analyzed the data with input from the other authors; J.M.M. wrote the paper with input from the other authors; J.U., K.F. critically reviewed the paper. All authors reviewed the results and approved the final version of the manuscript.

**Data Availability**
The data that support the findings of this study have been deposited in Zenodo at doi: 10.5281/zenodo.4926213; the data are available upon reasonable request.

**References**

1.  Nenadić, I.: Unpacking the "European approach" to tackling challenges of disinformation and political manipulation. Internet Policy Review. 8, (2019). https://doi.org/10.14763/2019.4.1436.
2.  Tenove, C.: Protecting Democracy from Disinformation: Normative Threats and Policy Responses. The International Journal of Press/Politics. 25, 517–537 (2020). https://doi.org/10.1177/1940161220918740.

3. Bennett, W.L., Livingston, S.: The disinformation order: Disruptive communication and the decline of democratic institutions. European Journal of Communication. 33, 122–139 (2018). https://doi.org/10.1177/0267323118760317.
4. Shao, C., Hui, P.-M., Wang, L., Jiang, X., Flammini, A., Menczer, F., Ciampaglia, G.L.: Anatomy of an online misinformation network. PLoS ONE. 13, e0196087 (2018). https://doi.org/10.1371/journal.pone.0196087.
5. Donovan, J.: Navigating the Tech Stack: When, Where and How Should We Moderate Content?, https://www.cigionline.org/articles/navigating-tech-stack-when-where-and-how-should-we-moderate-content.
6. EURid: APEWS, https://eurid.eu/en/register-a-eu-domain/apews/, last accessed 2021/03/30.
7. Floridi, L.: Trump, Parler, and Regulating the Infosphere as Our Commons. Philos. Technol. 34, 1–5 (2021). https://doi.org/10.1007/s13347-021-00446-7.
8. Au, Y., Howard, P.N., Ainita, P.: Profiting from the Pandemic. Oxford Internet Institute, Oxford (2020).
9. Guess, A.M., Nyhan, B., Reifler, J.: Exposure to untrustworthy websites in the 2016 US election. Nat Hum Behav. 4, 472–480 (2020). https://doi.org/10.1038/s41562-020-0833-x.
10. Klingova, K.: What Do We Know About Disinformation Websites in the Czech Republic and Slovakia?, https://www.globsec.org/news/what-do-we-know-about-disinformation-websites-in-the-czech-republic-and-slovakia/, last accessed 2021/03/23.
11. Mintal, J.M.: SLOVAKIA: SNAKE OIL SPILLS ONTO THE WEB. In: The Unbearable Ease of Misinformation. Center for Media, Data and Society at Central European University, Budapest (2020).
12. Konspiratori.sk: List of websites with controversial content, https://www.konspiratori.sk/, last accessed 2021/03/23.
13. Syrovátka, J., Vinklová, J., Zikmundová, A., Wojtula, L.: Disinformation Business Model. Prague Security Studies Institute, Prague (2020).
14. Krátka Špalková, V.: VÝROČNÍ ZPRÁVA O STAVU ČESKÉ DEZINFORMAČNÍ SCÉNY ZA ROK 2019 — Kremlin Watch Report. European Values, Prague (2021).
15. Konspiratori.sk: Criteria for including a website in the database, https://www.konspiratori.sk/en/inclusion-criteria.php, last accessed 2021/03/26.
16. WHOIS Search, Domain Name, Website, and IP Tools - Who.is, https://who.is/, last accessed 2021/03/23.
17. DNS Lookup - Check DNS All Records, https://dnschecker.org/all-dns-records-of-domain.php, last accessed 2021/03/23.
18. Domain to IP. Find Location of Websites or Domains, https://www.bulkseotools.com/bulk-domain-to-location.php, last accessed 2021/03/23.
19. ToolTrackerTracker < Dmi < Foswiki, https://wiki.digitalmethods.net/Dmi/ToolTrackerTracker, last accessed 2021/03/23.
20. WhatRuns – Discover What Runs a Website, https://www.whatruns.com/, last accessed 2021/03/23.

21. Research Solution & Data Analysis Tools | Similarweb, https://www.similarweb.com/corp/research/, last accessed 2021/03/23.
22. Abusive and Offensive Online Communications: A Scoping Report. Law Commission, London (2018).
23. Fabian, K., Mintal, J.M.: Cognitive Analysis of Security Threats on Social Networking Services: Slovakia in need of stronger action. (2020).
24. The Slovak Spectator: Health Ministry moves against disinformation website - spectator.sme.sk, https://spectator.sme.sk/c/22486133/health-ministry-moves-against-disinformation-website.html, last accessed 2021/04/17.
25. Klezl, T.: Dezinformací o vakcíně přibývá. Poplašnou zprávu o mrtvých na Gibraltaru řeší policie | Aktuálně.cz, https://zpravy.aktualne.cz/domaci/dezinformace-vakcina-covid/r~603aecd0647411eb95caac1f6b220ee8/, last accessed 2021/04/18.
26. Digital Citizens Alliance: DOMAINS OF DANGER : How website operators and Registrars trade Internet Safety for Profit. (2020).